\documentclass[usenatbib]{mn2e}
\usepackage{natbib,psfig}

\catcode`\@=11
\def\gsim{\ifmmode{\,\mathrel{\mathpalette\@versim>\,}}
    \else{$\,\mathrel{\mathpalette\@versim>}\,$}\fi}
\def\lsim{\ifmmode{\,\mathrel{\mathpalette\@versim<\,}}
    \else{$\,\mathrel{\mathpalette\@versim<}\,$}\fi}
\def\@versim#1#2{\lower 2.9truept \vbox{\baselineskip 0pt \lineskip
    0.5truept \ialign{$\m@th#1\hfil##\hfil$\crcr#2\crcr\sim\crcr}}}
\catcode`\@=12  

\def\Mdef{{M}_{\rm def}}

\def\Macc{{M}_{\rm acc}}

\def\d{{\rm d}}\def\e{{\rm e}}
\def\LB{L_{\rm B}}
\def\LV{L_{\rm V}}
\def\LEdd{L_{\rm Edd}}
\def\LHXnuc{L_{\rm HX,nuc}}
\def\Loptnuc{L_{\rm opt,nuc}}
\def\Lfivenuc{L_{\rm 5 GHz,nuc}}
\def\LX{L_{\rm X}}
\def\LBnuc{L_{\rm B,nuc}}
\def\LBsun{L_{\rm B\odot}}
\def\LVsun{L_{\rm V\odot}}

\def\Mstar{{M}_{*}}
\def\Lstar{{L}_{*}}

\def\Mbh{{M}_{\rm bh}}

\def\Hz{{\rm Hz}}

\def\Mdot{\dot{M}}
\def\Mdotsat{\dot{M}_{\rm sat}}

\def\kb{k_{\rm B}}
\def\phisat{\phi_{\rm sat}}
\def\uz{u_0}
\def\mp{m_{\rm p}}
\def\M{{M}}
\def\Msun{{M}_{\odot}}
\let\msun=\Msun

\def\sigmaz{\sigma_0}

\def\sigmazsat{\sigma_{\rm 0,sat}}
\def\sigmatilzsat{\tilde{\sigma}_{\rm 0,sat}}
\def\Stot{S_{\rm tot}}
\def\Ssat{S_{\rm sat}}
\def\dSsqvec{d^2{\bf S}}
\def\vsat{v_{\rm sat}}
\def\Temp{T}
\def\Tz{T_0}
\def\Tismseven{\left({\Tism \over 10^7 \kelvin}\right)}
\def\tanh{\mathop{\rm tanh}}

\def\arctanh{{\rm arctanh}}
\def\Tc{T_{\rm c}}

\def\tcool{t_{\rm cool}}
\def\tev{t_{\rm ev}}
\def\tevcyl{t_{\rm ev,cyl}}
\def\tdyn{t_{\rm dyn}}

\def\rmax{r_{\rm max}}
\def\rkpc{\left({r \over \kpc }\right)}

\def\Mdyn{{M}_{\rm dyn}}
\def\Mdyntot{{M}_{\rm dyn,tot}}
\def\Mgal{{M}_{\rm gal}}

\def\Mten{\left[{\Mdyn(r) \over 10^{10} \Msun }\right]}
\def\rhodyn{\bar{\rho}_{\rm dyn}}
\def\rhoc{\rho_{\rm c}}
\def\rhoism{\rho_{\rm ism}}
\def\lcrit{l_{\rm crit}}
\def\Mrad{{M}_{\rm rad}}

\def\Mcrit{{M}_{\rm crit}}
\def\arad{a_{\rm rad}}
\def\asat{a_{\rm sat}}
\def\Msat{{M}_{\rm sat}}

\def\Mmin{{M}_{\rm min}}
\def\asf{a_{\rm sf}}
\def\Msf{{M}_{\rm sf}}

\def\Tc{T_{\rm c}}
\def\Tism{T_{\rm ism}}
\def\csism{c_{\rm s,ism}}
\def\cs{c_{\rm s}}
\def\ne{n_{\rm e}}

\def\neismone{\left({\neism \over \cmmcube}\right)}
\def\nHcone{\left({\nHc \over \cmmcube}\right)}

\def\ne{n_{\rm e}}

\def\nHc{n_{\rm H,c}}

\def\neism{n_{{\rm e},{\rm ism}}}
\def\nez{n_{{\rm e},0}}
\def\press{P}
\def\pressism{P_{\rm ism}}
\def\qsat{q_{\rm sat}}
\def\qcl{q_{\rm cl}}
\def\qclvec{{\bf q}_{\rm cl}}

\def\cmmcube{\,{\rm cm}^{-3}}
\def\cm{\,{\rm cm}}

\def\Myr{\,{\rm Myr}}
\def\yr{\,{\rm yr}}
\def\Gyr{\,{\rm Gyr}}
\def\sm1{\,{\rm s}^{-1}}
\def\kelvin{\,{\rm K}} \let\K=\kelvin
\def\kpc{\,{\rm kpc}}
\def\akpc{\left({a \over \kpc }\right)}
\def\log{\,{\rm log}}
\def\ergs{\,{\rm erg}}
\def\kspitzer{\kappa_{\rm 0}}
\newcommand{\ls}{{_<\atop^{\sim}}}

\title[Thermal evaporation and galaxy formation]{The role of thermal evaporation in galaxy formation}

\author[C. Nipoti and J. Binney]{Carlo Nipoti$^1$\thanks{E-mail: carlo.nipoti@unibo.it} and James  Binney$^2$\\
$^{1}$Dipartimento di Astronomia, Universit\`a di Bologna, via Ranzani 1, 40127 Bologna, Italy\\
$^{2}$ Rudolf Peierls Centre for Theoretical Physics, Keble Road, Oxford OX1 3NP, UK\\}

\begin{document}

\date{Accepted 2007 September 18. Received 2007 September 18; in original form 2007 July 27}

\pagerange{\pageref{firstpage}--\pageref{lastpage}} \pubyear{2007}

\maketitle

\label{firstpage}

\begin{abstract}

  In colour-magnitude diagrams most galaxies fall in either the
  ``blue cloud'' or the ``red sequence'', with the red sequence
  extending to significantly brighter magnitudes than the blue
  cloud. The bright-end of the red sequence comprises elliptical
  galaxies with boxy isophotes and luminosity profiles with shallow
  central cores, while fainter elliptical galaxies have disky
  isophotes and power-law inner surface-brightness profiles. An analysis of
  published data reveals that the centres of galaxies with power-law central
  surface-brightness profiles have younger stellar populations than the
  centres of cored galaxies.

  We argue that thermal evaporation of cold gas by virial-temperature
  gas plays an important role in determining these
  phenomena.  In less massive galaxies, thermal evaporation is not
  very efficient, so significant amounts of cold gas can reach the
  galaxy centre and fill a central core with newly formed stars,
  consistent with the young stellar ages of the cusps of ellipticals
  with power-law surface-brightness profiles.  In more massive
  galaxies, cold gas is evaporated within a dynamical time, so during an
  accretion event star
  formation is inhibited, and a core in the stellar density
  profile produced by dissipationless dynamics cannot be refilled.  In
  this picture, the different observed properties of active galactic
  nuclei in higher-mass and lower-mass ellipticals are also explained
  because in the former the central supermassive black holes
  invariably accrete hot gas, while in the latter they typically
  accrete cold gas.

  An important consequence of our results is that at the present time
  there cannot be blue, star-forming galaxies in the most massive
  galactic halos, consistent with the observed truncation of the blue
  cloud at $\sim\Lstar$.
  
\end{abstract}


\section{Introduction}

Stars form from cold gas. Disk galaxies like the Milky Way or the Magellanic
clouds have significant quantities of cold gas in their disks, and this gas
has sustained star formation through most or all of the galaxies' lifetimes.
Elliptical galaxies lack cold gas, and for at least several gigayears these
systems have not formed significant numbers of stars. Photometry of galaxies
observed in the Sloan Digital Sky Survey shows that in a
colour-magnitude diagram galaxies predominantly lie in either a
``blue cloud'' or a ``red sequence'', with a smaller number of galaxies in a
``green valley'' between these features \citep{Blanton03,Baldry04,Driver06}.
Blue-cloud galaxies are forming stars, while red-sequence galaxies are not.
Both populations extend to faint magnitudes, but the red sequence extends to
significantly brighter magnitudes than the blue cloud.

Within the luminous elliptical galaxies of the red sequence two
sub-populations can be distinguished. At very bright magnitudes there
are objects with slightly boxy isophotes and luminosity profiles that
become shallow at small radii, while at fainter magnitudes the
galaxies mostly have disky isophotes and power-law inner
surface-brightness profiles
\citep[][]{Lau95,Fab97,Gra03,Tru04,Lau05,Fer06}.

In this paper we argue that all these phenomena are reflections of the way
gas at a galaxy's  virial temperature interacts with the cold gas required
for star formation.

During an episode of star formation, only a part of the gas reservoir
that drives the episode is converted into stars; some of the residual
gas is heated by supernovae to a temperature $\sim3\times10^6\K$
\citep{Larson,DekelS}. Dark halos with masses $\ls
\Mcrit\simeq10^{12}\msun$ have potential wells that are too shallow to
contain supernova-heated gas, so it flows out into intergalactic
space. More massive halos do confine the hot gas gravitationally, with
the result that once the halo's mass exceeds $\Mcrit$, the density of
hot gas in a halo builds up. As the density rises, the central cooling
time of the gas shortens, and the rate at which gas accretes onto the
central black hole (BH) increases.  Studies of cooling flows suggest
that the temperature sensitivity of the accretion rate onto the BH
leads to an unsteady equilibrium, in which jets powered by accretion
onto the BH replenish energy radiated by the gas
\citep{Birzan04,Bin05}. Hence halos with $\M>\Mcrit$ are filled with gas
at the virial temperature, and the density of this gas steadily increases.

We argue that such trapped hot gas eliminates cold gas by a
combination thermal conduction and ablation. In the more massive
galaxies of the red sequence, cold gas is eliminated within a
dynamical time, with the result that when an object that contains cold
gas falls into such a galaxy, the infalling gas has negligible chance
of reaching the galaxy's core before being heated to near the virial
temperature. Hence, once a core develops in the luminosity profile of
a massive galaxy of the red sequence, it cannot be be filled in by a
central burst of star formation. A corollary is that these systems do
not contain embedded disks, so their isophotes are more likely to be
boxy than disky.

In less luminous galaxies of the red sequence, the time required for
cold gas to be heated is likely to be longer than a dynamical time, so
when one of these galaxies encounters a gas-rich system, there is a
central burst of star formation that fills in the core in the
luminosity profile that is produced by dissipationless
dynamics. Moreover such central starbursts naturally account for the
fact that elliptical galaxies with power-law inner luminosity profiles
have younger central stellar populations (see Section~\ref{secobs}).

The galaxies of the blue cloud are expected to contain hot gas -- in
the case of the Milky Way, \cite{Spi56} already made this inference --
but such gas is typically not detected in X-ray observations,
suggesting that its density is lower than in red-sequence galaxies of
similar mass. Thus, evaporation of cold gas is not a significant
process in blue-cloud galaxies. However, in the last decade it has
become clear that star-forming disk galaxies cycle their interstellar
media through their halos several times over the Hubble time
\citep[e.g.][]{Fra06}. Consequently, once a potential well becomes so
deep that it confines supernova-heated gas, and gas accumulates at the
virial temperature, the cycling of the interstellar gas through the
halo makes the interstellar gas vulnerable to evaporation.  Star
formation ceases and the galaxy quickly moves from the blue cloud to
the red sequence.

The paper is organised as follows.  Section~\ref{seceva} defines the
relevant parameters and presents formulae from which they can be
calculated.  Numerical values for representative galaxy models are presented
in Section~\ref{secmod}.  When considering accretion events one has to
add the effects of encounters with systems that vary widely in mass; 
Section~\ref{secspec} provides a treatment of this problem. 
Sections~\ref{secegal} and
\ref{secblue} discuss the implications for elliptical galaxies
and galaxies of the blue cloud, respectively.  Our conclusions are in
Section~\ref{seccon}.

\section{Timescales for evaporation}
\label{seceva}

We want to determine the fate of a cloud of cool ($\Tc\ll 10^6 \kelvin$) gas
that falls through gas at the virial temperature $\Tism\sim10^6-10^7
\kelvin$. In particular, we ask whether a spheroidal cloud with semi-major
axis length $a$ that is on a
sufficiently low angular-momentum orbit can reach the galaxy centre and
form stars there.  

The motion of a cold cloud through a hot plasma is a complex dynamical
process, involving heat conduction, radiative cooling, ram-pressure
drag and ablation through the Kelvin-Helmholtz instability. We crudely
simplify the treatment of this process by assuming that the cloud
experiences only two opposed physical mechanisms: evaporation by
thermal conduction from the hot interstellar medium (ISM) and
condensation by radiative cooling.

The rate at which hot gas will ablate a cold cloud must depend on the
speed of the cloud's motion through the ambient gas, and will be
slowest for a stationary cloud; in this case a sheath of warm gas
builds up around the cloud, partially insulating it from the hot ambient
medium. In this paper we evaluate this minimum rate of ablation.  We
neglect the cloud's self-gravity, thus limiting ourselves to the case
of clouds much smaller than the host galaxy.

Depending on the physical properties of the ISM and on the size (and
geometry) of the cloud, the heat flux from the ISM to the cloud is
either classical \citep{Spi62} or saturated \citep{CowM77}.  For given
temperature and density of the ISM, we define the `saturation size'
$\asat$ such that the heat flux saturates for $a\lsim\asat$, and a
`critical size' $\arad$ at which radiative cooling balances heat
conduction: clouds bigger than $\arad$ condense the ambient medium,
while smaller clouds evaporate \citep[][and references
therein]{Mck77,NipB04}.  For clouds of size $a\lsim\arad$ we have to
compare the evaporation time $\tev$ with the dynamical time
$\tdyn$. Only clouds for which $\tev>\tdyn$ will survive long enough
to form stars.

\begin{table}
\begin{center}
\caption{List of symbols. The shape factor is $A(\epsilon)=1-\epsilon$
(oblate spheroidal clouds) and $A(\epsilon)=(1-\epsilon)^2$ (prolate
spheroidal clouds).\label{SymbTab}}
\begin{tabular}{ll}
$\asat(r)$&Conduction saturated for cloud size $a<\asat$\\
$\arad(r)$&Cloud evaporates for cloud size $a<\arad$\\
$\tev(r)$&Evaporation time for cloud size $a<\arad$\\
$\tdyn(r)$&Time to fall to the centre from $r$\\
$\asf(r)$&Minimum size of cloud that can reach centre from $r$\\
$\Mrad(r)$&Cloud mass for semi-major axis $\arad$: 
$\frac43\pi A(\epsilon)\rho_{\rm c}\arad^3$\\
$\Msat(r)$&Cloud mass for semi-major axis $\asat$: 
$\frac43\pi A(\epsilon)\rho_{\rm c}\asat^3$\\
$\Msf(r)$&Cloud mass for semi-major axis $\asf$: 
$\frac43\pi A(\epsilon)\rho_{\rm c}\asf^3$\\
$\Mmin$& $\max_{r>0}\Msf(r)$\\
\end{tabular}
\end{center}
\end{table}

Let us consider, for simplicity, a spherical galaxy for which we know
as functions of $r$, the electron temperature $\Tism(r)$ and density
$\neism(r)$ in the hot atmosphere. At each radius we can compute the
critical size $\arad(r)$, and the evaporation time $\tev(a,r)$ of a
cloud of size $a<\arad(r)$. At a given radius, the minimum size of a
cloud for it to survive evaporation and end up forming stars is the
``star-formation size'' $\asf(r)$ such that
$\tev(\asf,r)=\tdyn(r)$. Clouds bigger than $\asf$ can reach the
centre and there contribute to star formation. As the characteristic
sizes are strongly dependent on the cloud geometry, it is more
convenient to speak in terms of characteristic masses. Thus, for given
cloud shape and average mass density, we define the masses $\Mrad$,
$\Msat$, $\Msf$ of spheroidal clouds with semi-major axes $\arad$,
$\asat$, $\asf$, respectively.  To form stars in the central regions,
an infalling cloud must survive evaporation at all radii, so only
clouds more massive than $\Mmin=\max_{r>0}\Msf(r)$ can contribute to
central star formation. We will refer to $\Mmin$ as the ``minimum
cloud mass'', because in a given galaxy all clouds less massive than
$\Mmin$ will be evaporated and absorbed by the ISM within a dynamical
time.  Table~\ref{SymbTab} lists these definitions.

\subsection{Calculation of evaporation times}
\label{seccal}

Following \cite{CowS77}, we consider gas clouds modelled as prolate
and oblate spheroids, which can represent a wide range of geometries,
from filaments to disks, through spherical clouds. We consider oblate
and prolate spheroidal coordinates ($u,v$) related to the cylindrical
coordinates ($R,z$) by $R=\Delta \cosh u \sin v$, $z=\Delta \sinh u
\cos v$ (oblate), and $R=\Delta \sinh u \sin v$, $z=\Delta \cosh u
\cos v$ (prolate), where $\Delta$ is a length scale. The cloud is a
prolate or oblate spheroid with surface $u=\uz$.  The semi-major and
semi-minor axes of the cloud surface are\footnote{Note the different
  notation: here $a$ is the semi-major axis, while in \cite{CowS77}
  $a$ is the scale-length (our $\Delta$).}  $a=\Delta \cosh \uz$ and
$b=\Delta \sinh \uz$, and the cloud ellipticity is $\epsilon\equiv
1-b/a = 1-\tanh \uz$.  We assume that in the interface the electron
temperature and density are stratified with $u$.  At the cloud surface
($u=\uz$), the electron temperature is $\Tc\ll\Tism$, while
$\Temp\to\Tism$ for $u \to \infty$.

In the regime of unsaturated heat conduction ($a>\asat$), the heat
flux is given by the classical Spitzer formula
\begin{equation}
\label{eqqcl} 
\qclvec=-\kappa(T) {\bf \nabla} T,
\end{equation}
where the thermal conductivity is
\begin{equation}
\label{eqkappa2} 
\kappa(T)=f\kspitzer T^{5/2},
\end{equation} 
where $\kspitzer\simeq{1.84\times10^{-5}(\ln{\Lambda})}^{-1} \ergs
\sm1 \cm^{-1} \kelvin^{-7/2}$ \citep{Spi62}, $f\leq 1$ is the factor
by which magnetic fields suppress thermal conduction
\citep[e.g.][]{Bin81,Boh89,Tri89}, and $\ln\Lambda$ is the Coulomb
logarithm, which is only weakly dependent on $\ne$ and $T$ (in the
following we assume $\ln\Lambda=30$).  In the regime of classical heat
conduction and negligible radiative cooling ($\asat<a<\arad$),
\cite{CowS77} computed analytically the evaporation rate of prolate
and oblate clouds of cold ($\Tc\sim0$) gas immersed in a medium of
temperature $\Tism$: the evaporative mass loss rate is
\begin{equation}
\label{eqmdot}
\Mdot={16 \pi f  \mu \mp \kspitzer \Tism^{5/2} \Delta\over 25 \kb B(\epsilon)},
\end{equation}
 where $\mu$ is the mean gas particle mass in units of the proton mass
$\mp$, $\kb$ is the Boltzmann constant,
$B(\epsilon)=\frac{\pi}{2}-2\arctan[\tanh(\uz/2)]$ (oblate) and
$B(\epsilon)=\left|\ln [\tanh(\uz/2)]\right|$ (prolate), with $\uz=\arctanh
(1-\epsilon)$. In the spherical limit $\epsilon\to0$ ($\uz \gg 1$),
equation~(\ref{eqmdot}) yields the mass loss rate of a spherical cloud with
radius $a$ \citep{CowM77}.  Given the average mass density of clouds
$\rhoc$, the mass of a cloud of semi-major axis $a$ is
 \begin{eqnarray}
\label{eqm}
\M&=&{4\over 3} \pi A(\epsilon) a^3 \rhoc\simeq\\&\simeq&1.3\times 10^8 A(\epsilon) \akpc^3 \nHcone \Msun, 
\nonumber
\end{eqnarray}
where $A(\epsilon)=1-\epsilon$ (oblate) or
$A(\epsilon)=(1-\epsilon)^2$ (prolate), $\nHc$ is the cloud hydrogen
density, and when deriving the numerical value we have used $\rhoc\simeq
1.3 \mp\nHc$ (appropriate for abundances $Y=0.25$, $X=0.75$).
Combining equations~(\ref{eqmdot}) and (\ref{eqm}), we get the
evaporation time
\begin{eqnarray}\label{eqtev}
  &&\tev(a,r)\equiv{\M \over \Mdot}={25 
 \kb \rhoc a^2 C(\epsilon) \over 12 f \kspitzer \mu \mp
    \Tism^{5/2}}\simeq\\ &&\simeq 9.9\times 10^2 {C(\epsilon)
    \over f} \nHcone \Tismseven^{-5/2} \akpc^2 \Myr, \nonumber
\end{eqnarray}
where $C(\epsilon)\equiv A(\epsilon)B(\epsilon)\cosh[\arctanh
(1-\epsilon)] $, and we have used $\mu=0.59$.  The time for a cloud to
fall to the centre from $r$ is the dynamical time\footnote{The actual
  time will be somewhat longer because of the drag force (See
  Appendix~\ref{appdrag} for a discussion).}
\begin{eqnarray}
\label{eqtdyn}
\tdyn(r)&=&\sqrt{3 \pi \over 16 G \rhodyn}\simeq\\
&\simeq&7.44 \rkpc^{3/2}\Mten^{-1/2} \Myr, \nonumber
\end{eqnarray}
where $\rhodyn(r)=3\Mdyn(r)/ 4 \pi r^3$ is the average mass density
within $r$ \citep[e.g.][]{Bin87}.  From equation~(\ref{eqtev}) and
from the condition $\tev(\asf,r)=\tdyn(r)$, we derive the
star-formation size
\begin{equation}
\label{eqasf}
\asf\simeq0.32\left({f \over C(\epsilon)}
{\tdyn\over10^8\yr}\right)^{1/2}\! \nHcone^{-1/2}\!\Tismseven^{5/4}\!\kpc.
\end{equation}

As pointed out above, equation~(\ref{eqasf}) holds only if
$\asat<\asf<\arad$.  Postponing to Appendix~\ref{appeva} a detailed
treatment of the effects of saturation and radiation in the
evaporation of spheroidal clouds, we report here the equations
relating $\asat$ and $\arad$ to the physical properties of the ISM.
The saturation size is \citep{CowM77,CowS77}
\begin{equation}
\label{eqasat}
\asat= 5.0\times 10^{-4}{f^{1/2}\cosh{\uz}\over
  \sigmatilzsat\phisat}\Tismseven^2\neismone^{-1}\kpc,
\end{equation}
 where $\phisat\sim1$ and $\sigmatilzsat(\epsilon)$ are dimensionless
parameters ($0.05\lsim\sigmatilzsat \lsim 1$ for
$0.99\gsim\epsilon\gsim 0$, see Appendix~\ref{appsat}).  In the case
of significant saturation ($a<\asat$), we should solve the equations
for saturated heat conduction.  As pointed out by \cite{CowS77}, the
only analytic solution is for spherical symmetry (see
equation~\ref{eqmdotsat}), while in prolate or oblate geometry a
two-dimensional partial differential equation must be solved
numerically.  However, when $a<\asat$ the evaporation time-scale
computed for unsaturated thermal conduction provides a lower limit
(and the corresponding star-formation size $\asf$ an upper limit), and
we shall see that these limits together with the analytic solution
for the spherical case provide sufficient constraints for our
purposes.

In spherical symmetry, the critical size at which the cloud is
radiatively stabilised is \citep{Mck77}
\begin{equation}
\label{eqaradsph}
\arad= 1.9\times 10^{-2}{f^{1/2}\over\phisat}\Tismseven^2\neismone^{-1}\kpc.
\end{equation} 
Out of spherical symmetry the critical size cannot be computed
analytically. However, we assume $\arad$ as derived above to be
approximately correct also for oblate and prolate spheroids (see
Appendix~\ref{apprad} for a discussion).

Summarising, at each radius in a galaxy we have three characteristic
cloud masses: the star-formation mass $\Msf$, the saturation mass
$\Msat$, and the mass of radiatively stabilised clouds $\Mrad$. For
given $\epsilon$ and $\nHc$ these masses are computed from
equation~(\ref{eqm}), using the values of $\asf$, $\asat$ and $\arad$
from equations (\ref{eqasf}), (\ref{eqasat}) and (\ref{eqaradsph}).
Provided that $\Msat<\Msf<\Mrad$, the maximum of $\Msf$ over all radii
represents the minimum mass $\Mmin$ of a cool cloud that will survive
evaporation long enough to contribute to a central starburst.  If at
any radius $\Msf>\Mrad$, our neglect of cooling is invalid; then we
must assume $\Msf=\Mrad$, because clouds more massive than $\Mrad$
condense the ambient medium. On the other hand, if $\Msf<\Msat$,
$\Msf$ is just an upper limit to the actual value of the
star-formation mass.  At a given radius $r$, the characteristic masses
$\Msf$, $\Msat$ and $\Mrad$ scale with the suppression factor as
$f^{3/2}$, and depend on $\epsilon$ and $\nHc$.  The value of $\nHc$
can be constrained because we expect that the cloud is either in
pressure equilibrium with the ambient medium or overpressured (either
by gravity or supernova-driven expansion).  In particular, we consider
here clouds at $\Tc\sim10^4\kelvin$ with hydrogen density
$\nHc=\max[0.86\neism(\Tism/10^4\kelvin), 1 \cmmcube]$.

\begin{figure}
\centerline{\psfig{file=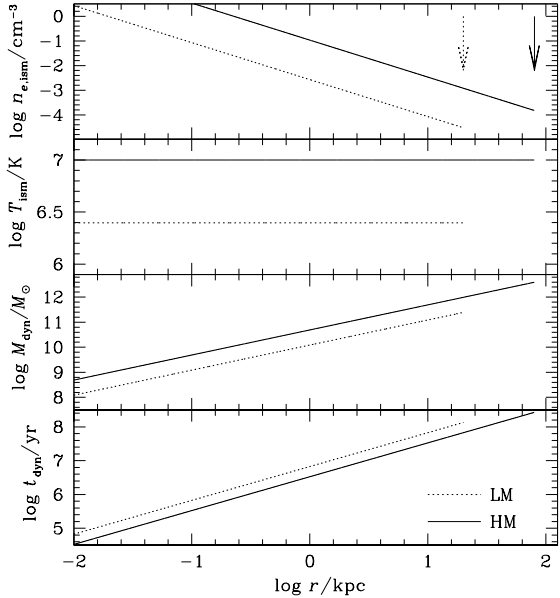,width=\hsize}}
\caption{From top to bottom: electron density and temperature of the
  ISM, dynamical mass and dynamical time as functions of radius for
  the galaxy models LM (dotted lines) and HM (solid lines). The arrows
  indicate the bounding radii $\rmax$ of the two models.}
\label{figmodel}
\end{figure}

\begin{figure*}
\centerline{\psfig{file=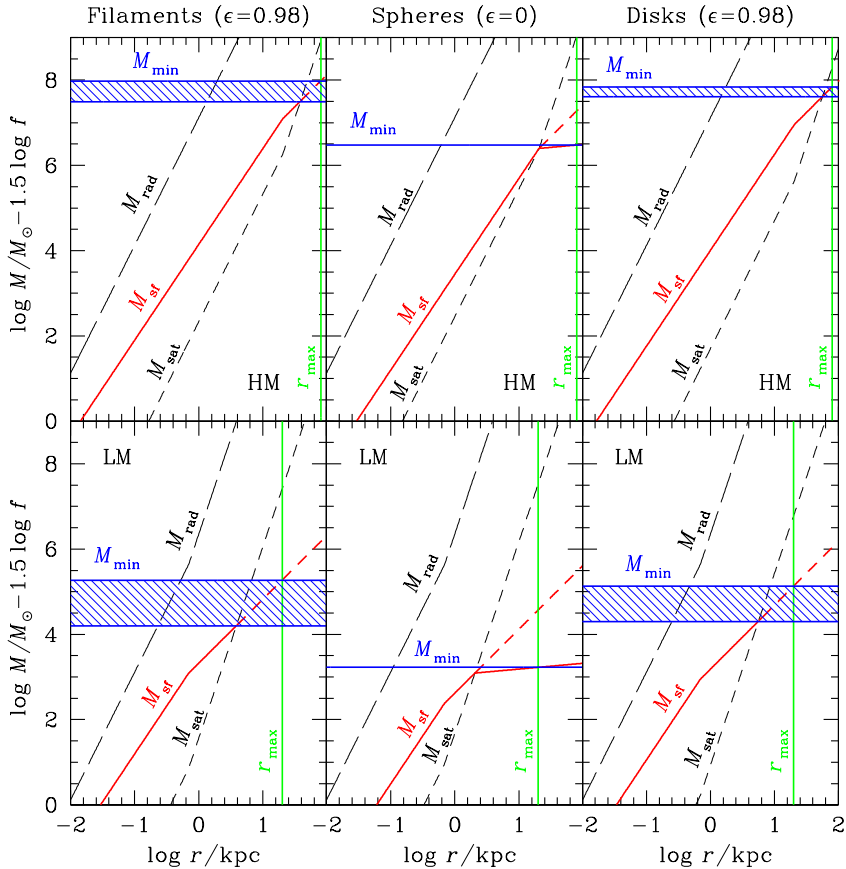,width=0.7\hsize}}
\caption{Characteristic cloud masses for models HM (top) and LM
  (bottom), for gas filaments (left), spherical clouds (centre), and
  gas disks (right). In each panel, the minimum mass $\Mmin$ (blue) is
  the maximum of $\Msf$ (red) over all radii smaller than $\rmax$
  (green); the dashed red line represents an upper limit of $\Msf$;
  the blue shaded region represents values of $M$ between the lower
  and upper limit of $\Mmin$; the black curves correspond to $M=\Mrad$
  (long dashed), and $M=\Msat$ (short dashed). Masses are in units of
  $f^{3/2}\Msun$, where $f$ is the factor by which thermal conduction
  is reduced below the Spitzer value.}
\label{figrm}
\end{figure*}

\section{Cloud masses in ellipticals}
\label{secmod}

Many elliptical galaxies have diffuse soft X-ray emission from a hot
atmosphere. In this section we investigate how such an atmosphere
affects the fate of any cold gas that might fall into the galaxy, for
example during a merger.
\begin{table}
\centering
\caption{ Parameters of the model galaxies LM (low mass) and HM
(high mass). $\Mgal$: baryonic mass. $\Mdyntot$: total dynamical mass.
$\Tz$: electron temperature of the ISM. $\nez$: electron number
density of the ISM at $r=1\kpc$. $\rmax$: bounding radius.
\label{GalTab}}
  \begin{tabular}{cccccc}
Model & $\Mgal$ & $\Mdyntot$ & $\Tz$ & $\nez$ & $\rmax$ \\
    & $\Msun$ & $\Msun$ & $\kelvin$ & $\cmmcube$ & $\kpc$ \\
\hline
LM &  $3.0\times10^{10}$ &  $2.5\times10^{11}$ & $2.5\times10^6$ & $2.7\times10^{-3}$ & $20$ \\
HM &  $3.0\times10^{11}$ &  $3.9\times10^{12}$ & $1.0\times10^7$ & $1.1\times10^{-1}$ & $80$ \\
\hline
\end{tabular}
\end{table}

Detailed radial temperature and density profiles are available for
only a few, mostly exceptionally luminous galaxies
\citep{Irw96,MatB03,Hum06,Fuk06,Dav06,Ran06,Osu07}.  While it is
straightforward to evaluate the parameters of Table~\ref{SymbTab} for
specific galaxies that have observationally estimated profiles, it is
more instructive to use the data to build representative model
galaxies and evaluate the quantities of Table \ref{SymbTab} for these
models.

Our models are spherical and within some bounding radius $\rmax$ have
perfectly flat circular-speed curves, so the total dynamical
mass interior to $r$ is $\Mdyn(r)/\Mdyntot=r/\rmax$, where
$\Mdyntot\equiv\Mdyn(\rmax)$.  We take the ISM to be in hydrostatic
equilibrium at a constant temperature $\Tism(r)=\Tz$.  From these
assumptions it follows that the density profile of the gas is a power
law in $r$ with a slope that depends on $\Tz$; we adopted
 \[
\neism(r)=\cases{\nez(r/\kpc)^{-3/2}&$r<\rmax$\cr
0&otherwise.}
\]
We consider two models, the parameters of which are reported in
Table~\ref{GalTab}: model LM represents a relatively low-mass,
gas-poor elliptical galaxy such as NGC~3377 or NGC~3245, while model
HM represents a high-mass, gas-rich elliptical galaxy such as NGC~4472
or NGC~4649. $\Mdyn(r)$, $\neism(r)$, $\tdyn(r)$, and $\Tism(r)$ are
plotted in Fig.~\ref{figmodel} for models LM (dotted lines) and HM
(solid lines).

\begin{figure}
\centerline{\psfig{file=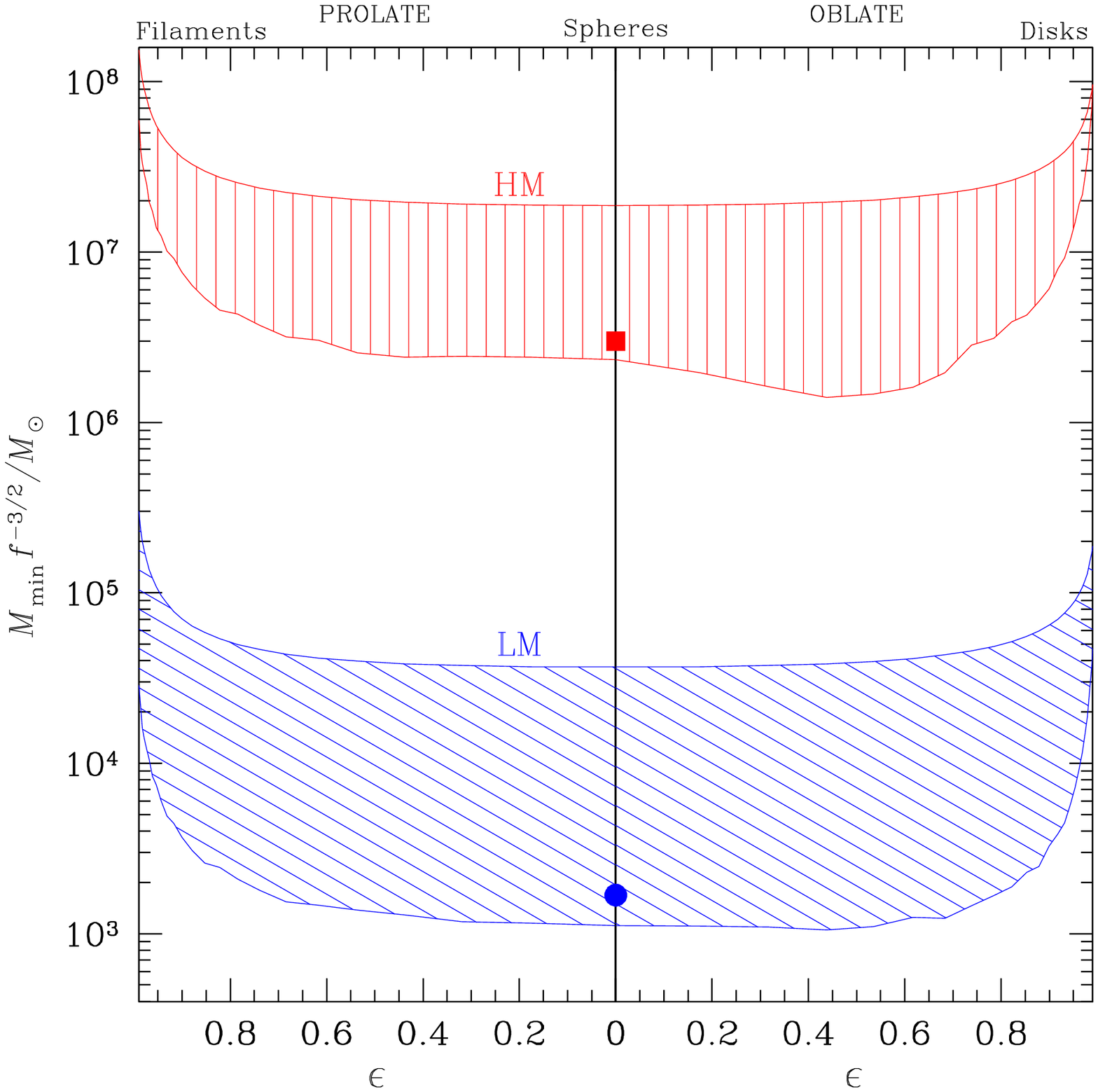,width=\hsize}}
\caption{Minimum mass as a function of $\epsilon$ for prolate (left)
  and oblate (right) clouds in units of $f^{3/2}\Msun$. The minimum
  mass lies in the region shaded with red vertical lines for model HM
  and in the region shaded with blue diagonal lines for model LM. The
  solid red square and blue circle indicate the values of $\Mmin$
  obtained for spherical clouds for model HM and LM, respectively.}
\label{figepsm}
\end{figure}

For models LM and HM we computed the characteristic masses $\Mrad(r)$,
$\Msf(r)$, $\Msat(r)$ and $\Mmin$ for oblate and prolate cloud models
with ellipticity in the range $0<\epsilon<0.99$.  In all cases (even
for thermal conduction suppression factor $f\sim 1$) we find
$\asf(r)\ll r$, where $\asf(r)$ is the size of a cloud of mass
$\Msf(r)$, consistent with our assumption of small
clouds. Figure~\ref{figrm} shows the characteristic masses (in units
of $f^{3/2}\Msun$) as functions of radius for models HM (top panels)
and LM (bottom panels) for three representative cloud shapes: gas
filaments (prolate cloud with $\epsilon=0.98$; left), spherical gas
clouds (centre), and gas disks (oblate cloud with $\epsilon=0.98$;
right). Figure~\ref{figepsm} plots $\Mmin$ as a function of $\epsilon$
for prolate and oblate clouds, for model HM (region shaded with
red vertical lines) and LM (region shaded with blue diagonal lines).
We checked these results by evaluating the parameters of the gas rich
galaxies NGC~4472 and NGC~4649 \citep{Irw96,Hum06,Ran06}, and the
gas-poor galaxies NGC~3377 and NGC~3245 \citep{Dav06}, finding results
similar to those of the models LM and HM, respectively.
Figs.~\ref{figrm} and \ref{figepsm} yield the following conclusions:

\begin{itemize}
\item[$\bullet$]The star-formation mass $\Msf$ is an increasing
  function of radius, partly because the dynamical time is and partly
  because of its dependence on $\nHc$. This indicates that cool clouds
  are more vulnerable in the outer regions of the galaxy. We note that
  in Fig.~\ref{figrm} the change of slope of the curves representing
  $\Mrad(r)$, $\Msat(r)$ and $\Msf(r)$ reflects the transition between
  overpressured clouds ($\nHc=1\cmmcube$; outer regions) and clouds in
  pressure equilibrium ($\nHc>1\cmmcube$; inner regions; see
  Section~\ref{seccal}).
\item[$\bullet$]In all cases $\Mrad$ is significantly higher than $\Msf$, so
  neglecting radiative cooling in the computation of $\asf$ is
  justified.
\item[$\bullet$]Saturation becomes important at large radii, where the
  density of the ISM becomes low. So $\Msf$ at large radii is just an
  upper limit (dashed red lines in Fig.~\ref{figrm}) and, as a
  consequence, only upper and lower limits on $\Mmin$ can be derived
  (shaded blue areas in Fig.~\ref{figrm}). An exception is the
  spherical case, for which we computed $\Msf$ also in the saturated
  regime, by using as evaporation time $\M/\Mdotsat$, with $\Mdotsat$
  given in equation~(\ref{eqmdotsat}). From Fig.~\ref{figrm} (central
  column of panels) it is apparent that $\Msf(r)$ for spherical clouds
  is almost constant in the saturated regime. As a consequence,
  $\Mmin$ (solid symbols in Fig.~\ref{figepsm}) lies close to the
  lower limit one would derive from the classical evaporation
  rate. This suggests that also in the case of prolate and oblate
  clouds the actual value of $\Mmin$ should be closer to the lower
  than to the upper limit. So, we expect the actual value of $\Mmin$
  to be basically independent of $\rmax$.
\item[$\bullet$]For given model galaxy, $\Mmin$ depends on the
  geometry of the cloud, with higher values obtained for very
  non-spherical systems (especially filaments) and the lowest values
  obtained in the case of spherical clouds (see Fig.~\ref{figepsm}).
  Thus for fixed mass, more non-spherical systems are more vulnerable
  to evaporation.
\item[$\bullet$]For fixed cloud geometry, $\Mmin$ is (in physical
  units) typically about three orders of magnitude higher in model HM
  than in model LM (Fig.~\ref{figepsm}).  More significantly, $\Mmin$
  normalised to the baryonic mass of the galaxy $\Mgal$ is a factor of
  $\sim 100$ higher in model HM than in model LM. Thus, proportionally
  bigger clouds are eliminated by evaporation in model HM than in
  model LM.
\item[$\bullet$]The actual value of $\Mmin$ in physical units is
  largely uncertain because of the dependence on $f^{3/2}$.  In the
  (unlikely) case of Spitzer conductivity ($f \sim 1$) we get
  $\Mmin\sim 10^7-10^8\Msun$ in model HM and $\Mmin\sim
  10^4-10^5\Msun$ in model LM, but for a more plausible value of the
  suppression factor, $f\sim 0.01$ \citep[][and references
  therein]{NipB04}, $\Mmin$ is as low as $\Mmin\sim 10^4-10^5\Msun$ in
  model HM and $\Mmin\sim 10-100\Msun$ in model LM.

\section{Aggregate gas infall}
\label{secspec}

The quantity $\Mmin$ is the minimum mass an individual cloud must have
to survive infall to the centre. To estimate the mass of cold gas that
becomes available for central star formation, we need to integrate
over the spectrum of cloud masses from this minimum mass to
infinity. We can only guess at the form of the spectrum. Clearly it
reflects the baryonic-mass spectrum of galaxies that are available to
be captured by the galaxy under study, which might be approximated by
the baryonic-mass function of all galaxies truncated at the baryonic
mass of the given galaxy to take account of the fact that an encounter
with a more massive galaxy leads to destruction rather than growth. In
reality this spectrum will overemphasise big clouds, both because
no galaxy has more than a small fraction of its mass in cold gas, and
because tidal shredding will split a galaxy into many clouds during a
merger. Therefore a reasonable hypothesis for the cloud spectrum is
the baryonic mass spectrum of galaxies first truncated at the mass of
the given galaxy and then scaled to smaller masses by a factor $\beta$
in the range $(0.01,0.1)$.

If we identify the baryonic-mass function with the luminosity function, and
approximate the latter with the \cite{Sch76} form, the mass of cold gas
that can reach the centre of a galaxy of baryonic mass $\Mgal$ is 
 \begin{equation}
\Macc\propto\Mstar\int_{\Mmin/\beta \Mstar}^{\Mgal/\Mstar}\d x\,x^{1-\alpha}\e^{-x},
\end{equation}
 where $\alpha\simeq1.25$ is the faint-end slope of the Schechter function,
and $\Mstar\simeq10^{11}\Msun$ is the baryonic mass of an $\Lstar$ galaxy. The
integrand above varies slowly below the exponential cutoff, so we can
approximate the integral by the between its lower limit and the smaller of
the upper limit and unity. Then the
dimensionless measure of the importance of cold accretion is
 \begin{equation}
{\Macc\over \Mgal}\propto\cases{
\min\left(1,{\Mstar\over\Mgal}\right)-{\Mmin\over\beta
\Mgal}&${\Mmin\over\beta\Mgal}<\min\left(1,{\Mstar\over\Mgal}\right)$\cr 0&otherwise.}
\end{equation}
Thus accretion shuts off completely for $\Mmin>\beta\min\left(\Mgal,\Mstar\right)$, and for
$\Mgal>\Mstar$ its importance declines faster than $\Mgal^{-1}$.

Since we require the mass function of gas-rich objects rather than the
luminosity function of galaxies, it is not clear that it is appropriate to
obtain $\alpha$ from the galaxy luminosity function: at early times the
spectrum of cloud masses should be the same as the  spectrum of
dark-matter halos, which has a steeper slope $\alpha\simeq2$. The flatter
spectrum of the galaxy luminosity function is thought to arise from the
difficulty of forming stars in low-mass halos, not from an absence of
low-mass gas clouds. If we take $\alpha=2$, the integral above should be
approximated by
 \begin{equation}
\int_{x_0}^{x_1}\d x\,x^{-1}\e^{-x}\simeq\cases{
\ln[\min(1,x_1)/x_0]&$x_0<\min(1,x_1)$\cr
0&otherwise}
\end{equation}
so
 \begin{equation}
{\Macc\over \Mgal}\propto\cases{
{\Mstar\over \Mgal}\ln\left[{\beta\Mgal\over\Mmin}\min\left(1,{\Mstar\over\Mgal}\right)\right]&${\Mmin\over\beta\Mgal}<\min\left(1,{\Mstar\over\Mgal}\right)$\cr 0&otherwise.}
 \end{equation}
 Again cold accretion declines in importance faster than $\Mgal^{-1}$
and cuts off for $\Mmin>\beta\min\left(\Mgal,\Mstar\right)$.

Figure~\ref{figmacc} shows the relative importance for the LM and HM
models of cold accretion from a spectrum of cloud masses. The full
curves apply if the latter has the same low-end slope as the mass
spectrum of cold dark matter (CDM) halos, while the dashed curves
apply if the low-end slope of the cloud spectrum is the same as that
of the galaxy luminosity function. In each case, two curves are
plotted, showing the extreme cases of only spherical clouds (thick
lines) and only filaments\footnote{In this case we used as $\Mmin$ the
lower limit found for $\epsilon=0.98$ prolate clouds.}  (thin
lines). When $1.5\log f-\log\beta<2$, the relative importance of
accretion in the LM and HM models is nearly independent of $f$, $\beta$
and geometry but increases with $\alpha$. Thus accretion will have the
biggest differential impact on LM and HM systems when the low-end
slope of the mass spectrum of clouds is steep. If $1.5\log
f-\log\beta>2$, either because conductivity is efficient or clouds
fragment strongly during mergers, accretion cuts off in HM systems and
the ratio plotted in Fig.~\ref{figmacc} shoots upwards. Accretion cuts
off for smaller values of $f$ is clouds are filamentary.

In summary, the importance of cold accretion, measured by
  $\Macc/\Mgal$, depends on the assumptions on the cloud mass spectrum
  and on the highly uncertain parameters $f$ and $\beta$. If $1.5 \log
  f-\log \beta \lsim 2$, $(\Macc/\Mgal)_{\rm LM}/(\Macc/\Mgal)_{\rm
  HM}$ will lie the range $3-30$.  If processes such as the
  Kelvin-Helmholtz instability are effective in breaking clouds into
  small fragments, $1.5 \log f-\log \beta$ could take on larger values
  and accretion would be suppressed in HM systems.
\end{itemize}

\begin{figure}
\centerline{\psfig{file=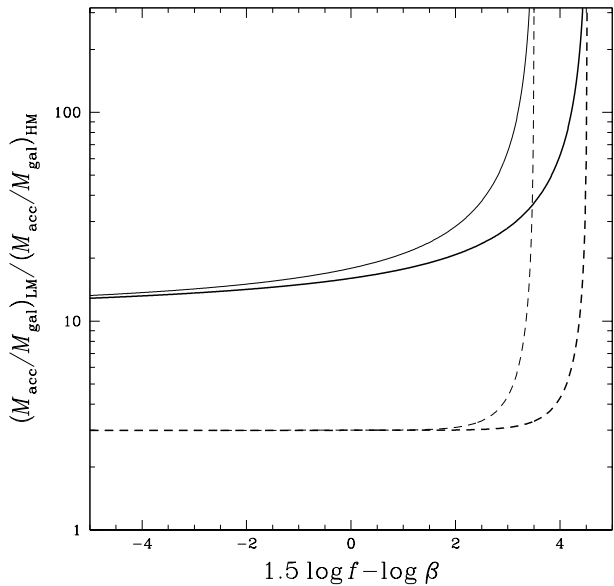,width=\hsize}}
\caption{The ratio between the quantity $\Macc/\Mgal$ in the LM
    model galaxy and the same quantity in the HM model galaxy as a
    function of the efficiency of thermal conduction $f$ and of the
    parameter $\beta$ defined in Section~\protect{\ref{secspec}}
    ($\Macc$ is the mass of accreted cold gas that can reach the
    galaxy centre and $\Mgal$ is the baryonic mass of the galaxy).
    The full curves apply if the cloud mass spectrum has the same
    low-end slope as the mass spectrum of CDM halos, while the dashed
    curves apply if the slope is that of the galaxy luminosity
    function. Thick and thin curves refer to spherical clouds and
    filaments, respectively.}
\label{figmacc}
\end{figure}

\section{Implications for ellipticals}
\label{secegal}

Figures~\ref{figepsm} and \ref{figmacc} show that in an X-ray luminous
galaxy a cold cloud has to be $\sim10^3$ times more massive to reach
the centre than in an elliptical with weak X-ray emission, and
aggregate gas in infall available for central star formation is more
important in the latter than in the former.  Thus we expect cold gas
and its associated star formation to have a much bigger impact on the
centres of X-ray weak ellipticals than on the centres of their X-ray
luminous brethren. In this section we discuss the nature of this
impact and relate it to a dichotomy in the observed properties of
elliptical galaxies.

\subsection{Observed properties of power-law and core galaxies}
\label{secobs}

On the basis of their central surface brightness (SB) profiles,
luminous elliptical galaxies fall into two distinct classes: power-law
(or Sersic) galaxies (PLGs), with SB profiles well represented by the
\cite{Ser68} law down to the centre, and core (or core-Sersic)
galaxies (CGs), with SB profiles deviating from the Sersic law in the
central regions because of the presence of a flat core
\citep[][]{Lau95,Fab97,Gra03,Tru04,Lau05,Fer06}.  Quantitatively, the
corresponding intrinsic stellar density profiles have inner
logarithmic slope $\gamma>0.5$ in PLGs, and $\gamma<0.3$ in CGs, with
few ``intermediate'' galaxies, having $0.3<\gamma<0.5$
\citep{Rav01,Res01,Lau07}.  It is useful to summarise the main
observed properties of PLGs and CGs.

\begin{figure}
\centerline{\psfig{file=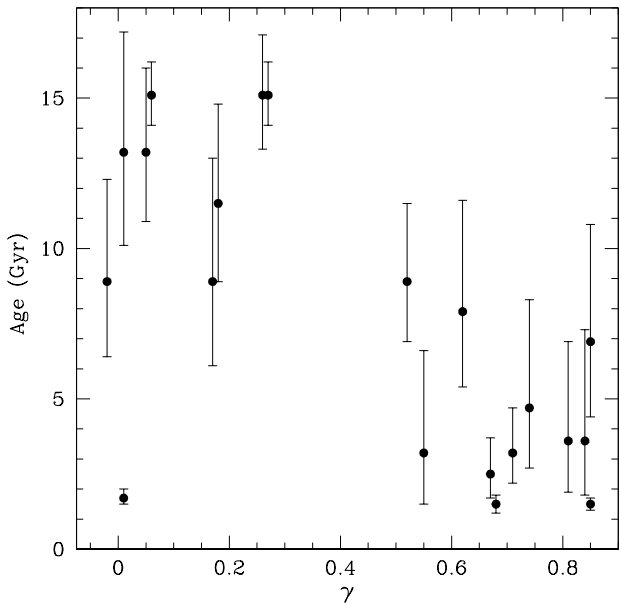,width=\hsize}}
\caption{The age of the central stellar population versus the central
slope of the SB profile $\gamma$ for a subsample of the elliptical and
lenticular galaxies studied by \protect\cite{Mcd06}. Central ages
(with error bars) are from \protect\cite{Mcd06}, while the values of
$\gamma$ are from \protect\cite{Lau07}.}
\label{figage}
\end{figure}

\begin{enumerate}
\item {\it Global galaxy properties:}
\begin{enumerate}

\item CGs are typically more luminous than PLGs, with a dividing
  luminosity $\LV\sim 3\times10^{10}\LVsun$. In particular, all
  galaxies brighter than $\LV\sim 5\times10^{10}\LVsun$ are CGs, while
  all galaxies fainter than $\LV \sim2\times 10^{10}\LVsun$ are PLGs.  At
  intermediate luminosities both types coexist \citep{Fab97,Lau07}.

\item CGs have boxy isophotes and rotate slowly, PLGs have disky
isophotes and rotate rapidly \citep{Kor96,Fab97}.

\item Almost all the brightest cluster galaxies are CGs \citep{Lai03}.

\item All galaxies with high X-ray emission from hot gas (high $\LX/\LB$)
are CGs, while ellipticals with lower $\LX/\LB$ include both CGs and PLGs
\citep{Pel99,Pel05,Ell06}.
\end{enumerate}

\item {\it Properties of the central regions:}
\begin{enumerate}
\item All ellipticals have central colour gradients, becoming redder
  towards the centre. CGs have weaker central colour gradients than do
  PLGs, though the difference is small and there is large scatter
  \citep{Lau05}.
\item CGs are rounder than PLGs at small radii.  PLGs typically have
  central stellar disks, while CGs do not \citep{Lau05}.
\item Nuclei (typically bluer than the surrounding galaxy) are more
  frequent in PLGs than in CGs \citep{Lau05}.  In the Virgo cluster no
  galaxies brighter than $\LB\sim 2.5\times 10^{10}\LBsun$ have nuclei
  \citep{Cot06,Fer06}.
\item Lower-velocity dispersion ellipticals have significantly younger
  central stellar populations than higher-velocity dispersion
  ellipticals \citep[][and references therein]{Mcd06}. This should
  imply a correlation between the stellar age of the inner regions of
  ellipticals and the central slope $\gamma$ of the SB
  profile. Combining information on the central age from \cite{Mcd06}
  and on $\gamma$ from \cite{Lau07}, we considered a sample of 20
  elliptical and lenticular galaxies, which turn out to be nicely
  segregated in a diagram plotting central age versus $\gamma$
  (Fig.~\ref{figage}), with CGs and PLGs having median central ages
  $13.2 \Gyr$ and $3.6 \Gyr$, respectively\footnote{The only outlier
  is NGC~4382, a core galaxy ($\gamma=0.01$) with an estimated central
  age $\sim 1.7 \Gyr$. We note that NGC~4382 has relatively low
  $\LX/\LB$ \citep{Pel05} and is characterised by atypical central SB
  profile and inner colour gradient \citep{Lau05}.}.
\end{enumerate}

\item {\it Activity of the central black hole:}
\begin{enumerate}
\item  The mean ratio
  $\Loptnuc/\LEdd$ of the optical nuclear emission and  
  the Eddington luminosity of the central BH
  is higher by two orders of magnitude in PLGs than in CGs \citep{Cap06}.

\item If $P_{1.4}$ is the global power of both core and extended
emission at $1.4\,$GHz CGs cover the full range ($18\lsim \log
P_{1.4}/W \Hz^{-1} \lsim26$), while PLGs show no significant radio
activity \citep{Der05}.

\item With $R=\Lfivenuc/\LBnuc$  the ratio of the radio to
B-band luminosities of the nucleus and $R_X=\Lfivenuc/\LHXnuc$ the
ratio of radio and $2-10\,$keV X-ray luminosities, CGs are radio-loud ($\log R \sim
3.6$; $\log R_X \sim -1.3$), while PLGs are radio-quiet ($\log R \sim 1.6$;
$\log R_X \sim-3.3$) \citep{Cap06,Balma06}.
\end{enumerate}
\end{enumerate}

\subsection{Formation mechanisms of the central stellar profiles}

During hierarchical galaxy formation luminous galaxies arise from
mergers of lower-luminosity systems. Lower-luminosity galaxies tend to
be PLGs, so the natural starting point for discussing the origin of
the CG/PLG dichotomy is the morphology of PLGs. That is, we assume
that at early enough times any elliptical galaxy is a PLG, and our
task is to explain why some systems became CGs. In this spirit one
associates with each CG a mass deficit $\Mdef$ equal to the mass that
would have to be added to the core to make the SB profile that of a
PLG: typically $\Mdef\sim\Mbh$, where $\Mbh$ is the central BH mass
\citep[][]{Mil02,Gra04,Mer06}. This observation suggests the following
popular ``core-scouring'' scenario, which has been validated by N-body
simulations \citep{Mil01}\footnote{Core scouring is not the only
mechanism able to produce cored profiles: \cite{Nip06} showed that
galaxies with SB profiles just like those of CGs can be obtained by a
straightforward dissipationless collapse in the absence of a central
BH.}: during dissipationless merging, the orbiting BHs reduce the
central star density by scattering stars out of the most tightly bound
orbits \citep{Beg80,Ebi91,Mak96,Fab97,Qui97,Vol03}.  Thus in the
core-scouring scenario the central slope $\gamma$ is determined by the
merging history of the galaxy, and the mass deficit $\Mdef$ is
naturally of the order of the masses of the participating BHs.  The
core-scouring scenario is successful in reproducing some of the
observations, but it cannot be a complete
explanation\footnote{\cite{Lau05} conclude ``Overall, the evidence is
consistent with the binary BH core formation mechanism, but it is not
a strong endorsement''.} because no purely stellar-dynamical mechanism
can introduce a characteristic mass scale and thus explain {why PLGs
are never luminous, while CGs often are.} Moreover, core scouring
alone does not explain the origin of the correlation between the
presence of cores and strong diffuse X-ray emission
\citep{Pel05,Ell06}, nor does it explain why flat cores have not
formed in PLGs, given that low-luminosity ellipticals such as M32 also
harbour BHs \citep{Fab97}.

\subsection{The role of thermal evaporation in determining the power-law/core
  dichotomy}

The discussion above suggests that a plausible working hypothesis is that
all ellipticals at some stage in their evolution have central cores.  The
results of Sections~\ref{secmod} and \ref{secspec} and the relative youth of
the central regions of PLGs (Fig.~\ref{figage}) suggest that in PLGs the
core created by dissipationless dynamics has been filled in by star
formation from cold gas that fell in during or after a merger. The aggregate
mass of new stars formed will be proportional to the quantity $\Macc$
introduced in Section~\ref{secspec}. The proposal requires that
$\Macc\ga\Mdef$ in PLGs and $\Macc<\Mdef$ in CGs. We know that
$\Mdef\sim\Mbh$, and $\Mbh$ is proportional to the total stellar mass of the
galaxy \citep{Mag98,Mar03}, so it is reasonable to assume
$\Mdef\propto\Mgal$.  So $\Macc/\Mdef\propto \Macc/\Mgal$, and our finding
that $\Macc/\Mgal$ is higher in hot-gas poor than in hot-gas rich
ellipticals (Fig.~\ref{figmacc}) implies that the former are more likely to
fill any central core.  Quantitatively, this result will depend on the
properties of the cloud mass spectrum.  In Section~\ref{secspec} we concluded
that when the power-law slope $\alpha$ of the mass function of clouds is
similar to the faint-end slope of the galaxy luminosity function, the mass
of gas that can reach the centre, normalised to the galaxy mass $\Mgal$ is
at least a factor of 3 higher in lower-mass ellipticals than in higher-mass
ellipticals. This difference rises to at least a factor of 15 if
$\alpha\sim2$, as it would be if the cloud mass spectrum were the same as
that of CDM halos. For very massive galaxies for which $\Mmin$ could be
almost as high as $\beta\Mstar$ very little cold gas can reach the galaxy
centre (see Section~\ref{secspec}).  So, we expect fainter galaxies to
accumulate cold gas that supports central star formation, while the more
massive galaxies can at most partially fill in the core created by
dissipationless dynamics. The hypothesis that since its last major merger or
accretion event every PLG has experienced an episode of central star
formation from cold, rapidly rotating gas naturally explains all the
observations listed in Section \ref{secobs} under the headings (i) and (ii):

\begin{itemize}
\item[(ia)] It is possible for a faint, gas-poor elliptical to be a CG
galaxy simply because it has not encountered a source of cold gas
since its last major merger. The brightest ellipticals have dense
virial-temperature atmospheres,
so no luminous galaxy should be a PLG.

\item[(ib)] The boxiness and slow rotation of CGs follows from the
dominant role played by dissipationless dynamics in their formation.

\item[(ic)] If a prerequisite for being a CG is the ability to confine hot
gas, these galaxies will be, or have been, the central galaxies of
groups and clusters.  When two groups merge, their central galaxies
will merge, thus ensuring that the central galaxy of the new group is also a
CG. 

\item[(id)] We have seen that all X-ray luminous galaxies should be the
central galaxies of groups or clusters. Such a galaxy can be a PLG only if
its dense atmosphere built up {\it after\/} its last major merger. This is
in principle possible but unlikely. When a small group falls into a rich cluster, it
is possible that the central galaxy will be stripped of its dark halo
before it is eaten by the central galaxy of the cluster. In this case
we will observe a CG that is not X-ray luminous. There should be PLGs
in all environments.

\item[(iia)] The late burst of star formation that filled in the
post-merger core of a PLG can be expected to be metal-rich and enhance
the galaxy's metallicity gradient.

\item[(iib)] The dynamical importance of rotation will increase as cold gas
streams in, facilitating the formation of a central disk and/or a rapidly
rotating kinematically decoupled core.

 \item[(iic)] If infall continues long after the merger \citep{Schweizer}
and cold gas can reach the centre, a blue nuclear cluster can form.

\item[(iid)] If stars cannot now form at the centres of CGs, the
relative youth of the central stellar populations of PLGs is explained.

\end{itemize}

\subsection{Mode of activity of the central black hole}
\label{secagn}

We now show that our picture also explains the connections between
galaxy morphology and active galactic nucleus (AGN) listed in Section
\ref{secobs} under the heading (iii).

There are two very different modes of BH accretion and feedback, usually
referred to as ``cold mode'' (or QSO mode) and ``hot mode'' (or radio-mode)
\citep[e.g.][]{Bes05,Bin05,Chu05,Har07,Sij07}. In short, these
are the main properties of the two modes:

\begin{enumerate}
\item[$\bullet$]{\it Cold mode}: the BH feeds from cold gas with
  $L\sim\LEdd$, with most of the energy released going into photons
  (optical, UV, X-ray).  The BH grows significantly in mass, and there
  is important attendant star formation.  This mode is dominant at
  high redshift and leads to the correlation between BH and spheroid
  mass. The most extreme manifestations of cold mode accretion are
  luminous QSOs.

\item[$\bullet$]{\it Hot mode}: the BH feeds from hot gas at
  $L\ll\LEdd$, and most of the energy released is mechanical, being
  associated with bipolar flows, which generate significant radio
  emission. The BH mass does not increase significantly, and star
  formation is inhibited. The classic hot-mode accretor is
  $\hbox{Virgo A}=\hbox{M87}$ \citep{Dimatteo03}, but FRI radio sources
  probably all belong to this class.
\end{enumerate}

Cold-mode accretors can be either radio-loud or radio-quiet, while hot-mode
accretors are invariably radio-loud.

In our picture all CGs must be hot-mode accretors, while PLGs can be
either hot- or cold-mode accretors, depending on whether they happen
to have encountered some cold gas recently. Items (iiia), (iiib) and
(iiic) of Section \ref{secobs} are consistent with this expectation, and taken together
suggest that most PLGs are in fact accreting cold gas.

A possible explanation of the radio-loudness of CGs is the suggestion
of \cite{Cap06} that a galaxy's merging history determines both the
presence of the core and the spin of the BH, and that the latter
controls radio-loudness \citep[see also][]{Wil95,Cen07,Sik07,Vol07}.
If BH spin were the only factor determining radio-loudness, the
radio-loudness of a BH could only change slowly with
time. Observations of micro-quasars show that a single source
---regardless of whether it is powered by a BH or by a neutron star---
often alternates short bursts of radio-loudness with longer periods of
radio quiescence \citep{NipBB05}. In addition, the long-term time
variability of central radio sources in galaxy clusters (once properly
scaled for the BH mass) is consistent with that of microquasars
\citep{NipB05}. These two pieces of evidence suggest that
radio-loudness is more likely to be controlled by the accretion mode
than by BH spin.  The results of the present paper show that the
observed radio properties of PLGs and CGs fit consistently in this
scenario.

Thus a considerable body of observational evidence points 
to a scenario in which the ability of
cold gas to survive evaporation from the hot ISM is responsible for
determining both the inner stellar profiles of elliptical galaxies and
the radio emission from their central supermassive BHs.

\section{End of the blue cloud}
\label{secblue}

As outlined in the Introduction, a robust bimodality is observed in
the properties of galaxies, which in colour-magnitude diagrams are
nicely segregated into a red sequence (extending to the brightest
magnitudes) and a blue cloud (truncated at $\sim \Lstar$). These
features, together with the shape of bright-end of the galaxy
luminosity function, are hard to reproduce in standard galaxy
formation models based on the proposal by \cite{ReeO77} and
\cite{WhiR78} that galaxies form by cooling virial-temperature gas.

Until recently it has been widely assumed that during virialization of a
dark-matter halo gas is heated to the virial temperature and that galaxies
form through the cooling of this hot gas. However, both analytic and
numerical results suggest that during virialization only a fraction of the
gas is heated to the virial temperature, and that this fraction is
negligible in halos less massive than a critical mass
$\Mcrit\approx10^{12}\msun$ \citep{Bin77,BirD03,Ker05,DekB06}. Moreover, as
\cite{Bin04} first pointed out, observations of cooling flows suggest that
heating by AGN-powered jets ensures that gas trapped at the virial
temperature never cools. Hence this gas is not available for star formation,
and galaxies can only go on forming stars and remain in the blue cloud to
the extent that they can acquire cool gas. Hence the process studied above,
of thermal evaporation of cold gas by hot, bears on the origin of the entire
Hubble sequence as well as the morphological dichotomy of elliptical
galaxies.

While there is now a wide measure of agreement that in very massive galaxies
star formation is ``quenched'' through lack of cold gas, the mechanism
responsible for cutting off the supply of cold gas is controversial.
\cite{DekB07} argue that gravitational heating alone suffices \citep[see also][]{KhochfarO}. While the
efficiency of gravitational heating undoubtedly rises with the clustering
mass scale, several observations indicate that non-gravitational heating is
important.  Indeed studies of the temperature-luminosity correlation of
diffuse X-ray sources has long provided strong evidence for powerful
non-gravitational heating \citep{Kaiser,PonmanSF}. Other indications are the
small fraction of baryons contained in galaxies, both in groups like the MW
and in rich clusters, the significant fraction of the products of
nucleosynthesis that are in intracluster media \citep{Renzini}, and the
direct observation of massive outflows from star-forming galaxies at both
small and high redshift \citep{Stricklandetal,Pettinietal}.

The availability of non-gravitational heating is crucial because gravity
alone will never drive gas {\it out\/} of dark matter halos, and the fact
that at faint magnitudes the galaxy luminosity function falls further and
further below the mass function of dark halos has long been attributed to the
ability of non-gravitational heating to drive gas out of low-mass halos
\citep{WhiR78,DekelS}.  So it seems clear that energy released during star
(and possibly black-hole) formation prevents virial temperature gas
accumulating in low-mass halos.  When cool gas falls into these systems it
is not heated to the virial temperature, so star formation can continue in
them.

It has been suggested that quasars are responsible for terminating
star formation in their hosts \citep{Spring05}. We feel this proposal
is implausible for two reasons: (i) The abundant evidence for (a) an
association between the masses of black holes and their host bulges,
(b) the coincidence between the peaks in the cosmic star-formation
rate and the quasar luminosity density, and (c) the rapidity of bulge
formation, indicates that quasars are associated with rapid star
formation, rather than quenching of star formation. (ii) Observations
of disk galaxies imply that cold infall is a sustained process, while
quasars flare up and then die. So while a quasar may clear its host of
cold gas, it cannot be responsible for keeping the host clean after
its death -- which is the great majority of the host's life. In our
view black holes have a vital role to play in quenching star
formation, but their role is an indirect one: acting as thermostats
for virial-temperature gas, which bears direct responsibility for
quenching star formation. The connections between quasars and bulges
arise because when cold gas is available on non-circular orbits (for
example during a merger), bulge stars form rapidly on non-circular
orbits and black holes accrete rapidly in an environment that causes
energy released by accretion to be rapidly degraded into low-energy
photons that produce only weak feedback on the interstellar medium
\citep{Sazonov}.

When a halo reaches the critical mass $\Mcrit$ two things happen
almost simultaneously: (i) the fraction of infalling gas that stays below
the virial temperature starts to dwindle, and (ii) the virial temperature
rises above the temperature to which star and black-hole formation heat gas,
so a hot atmosphere begins to accumulate -- the temperature of this
atmosphere is subsequently adjusted to the virial temperature by hot-mode
accretion onto the black hole. We do not yet know how precise is the
coincidence of the mass scales associated with the apparently independent
processes (i) and (ii). However, it is does seem likely that this mass scale
accounts for the upper limit to the luminosities of objects in the blue
cloud, and that thermal evaporation plays a significant role in setting this
limit. Semi-analytic models show that several features of the
distribution of galaxies in colour, magnitude and redshift can be accounted for if
there is a critical halo mass above which star-formation is quenched
\citep[e.g.][]{Cro06,Cat06,DekB06}.

A key object, that seems to be on the cusp of the transition across
$\Mcrit$ is the edge-on spiral galaxy NGC 5746. A {\it Chandra} observation
shows that this galaxy is enveloped by a nearly spherical halo of X-ray
emitting gas \citep{Rasmussen06}. The authors suggest that star formation in
the disk is sustained by cooling of this halo in the manner envisaged by
\cite{WhiR78}, but it is hard to understand why cooling and star formation
is not concentrated at the centre of the bulge, where the halo is densest
and the cooling time shortest, rather than further out in the disk, nor how
such a nearly spherical cloud could give rise to a disk in circular motion.
It seems much more likely that in NGC 5746 we see the first stages in the
growth of the body of trapped virial-temperature gas that will shortly
quench star formation, causing the galaxy to move onto the red sequence.

The red sequence extends to faint magnitudes because galaxies hosted by
halos that are much less massive than $\Mcrit$ fall into halos of mass
$M>\Mcrit$ and then quickly become red because (i) their own cold gas is
stripped out of them by a combination of ablation, thermal conduction and
supernova-driven flows, and (ii) their prospects of accreting fresh cold gas
are small given that they are now orbiting inside a halo of mass $M>\Mcrit$
and moving too fast to merge with other galaxies, which anyway have little
cold gas to offer. \cite{Postmanetal} showed that the density-morphology
relation is driven by the growth with density in the number of S0 galaxies
at the expense of spirals, and \cite{Bedregal06} and \cite{Barr07} have
recently proven beyond reasonable doubt that spiral galaxies passively fade into S0
galaxies, presumably as a result of gas-starvation. Another indication
that many red-sequence galaxies orbit in more massive halos is the fact that
they are preferentially found in dense environments, while blue-cloud
galaxies typically reside in lower-density environments
\citep[e.g.][]{Baldry06}.

In summary, a galaxy will move from the blue cloud to the red sequence
{\it either\/} because it sits in a primary halo and the mass of this
has just surpassed $\Mcrit$, {\it or\/} because it starts to orbit
sufficiently deeply in the potential of a primary halo with mass
$>\Mcrit$.  From this hypothesis it should be possible to predict the
luminosity functions of the blue cloud and the red sequence as
functions of redshift, and the clustering properties of red-sequence
galaxies. However, to make these predictions one needs both a detailed
picture of the clustering of dark halos and the distribution of
virial-temperature gas. The former could be taken from the halo model
\citep{Seljak00}, but there is currently no obvious source, either
theoretical or observational, from which to draw the latter:
determining the distribution of hot gas observationally is hard
because extended, low-density virial-temperature gas is a weak X-ray
emitter and current X-ray detectors do not have useful velocity
resolution. Building a model of the hot gas from cosmological
simulations that include baryons is problematic because of the
importance for gas of feedback from stars and AGN.

\section{Summary and conclusions}
\label{seccon}

Cool gas is a fundamental ingredient for galaxy formation and
evolution.  Naturally, the study of cool gas in galaxies has been
mainly focused on late-type galaxies, in which cool gas is most
abundant. Early-type galaxies are rather poor in cool gas, but thanks
to recent surveys, in CO \citep{Sage07}, HI \citep{Mor06,Oos07} and
optical lines \citep{Sar06}, we now have a detailed picture of the
properties of the cool gas also in these systems. In elliptical
galaxies such cool gas is immersed in a sea of plasma at the galaxy's
virial temperature, and a long-standing question is how these two
phases coexist and interact. In previous works evaporation of cool gas
by thermal conduction from gas in the hot phase has been invoked in
order to explain the shutdown of star formation in massive galaxies
\citep{Bin04,DekB06} and the observed ionised-gas kinematics relative
to that of the stars in early-type galaxies \citep{Sar07}, but
quantitative estimates of the importance of this process were lacking.

This paper gives a quantitative study of the evaporation of cold gas
clouds by thermal conduction from the hot ISM. We focused on two
classes of luminous elliptical galaxies, the low-mass (LM) and
high-mass (HM) systems, both of which probably have potential wells
deep enough to trap gas heated by star formation, but which differ in
the densities of their virial-temperature atmospheres. We have shown
that in the LM systems accreted cold gas is much more likely to feed
central star formation than in the HM systems.  This fact naturally
explains the observed dichotomy between PLGs and CGs in central
surface brightness profiles, X-ray luminosity, and activity of the
central supermassive BH.

Our exploration is based on a rather simplified analytical model of the
interaction between cold and hot gas, in which we account in detail only for
thermal conduction and radiative cooling because we expect these to be the
dominant physical processes.  Ideally, the next step would be to explore the
problem of cold-gas accretion and evaporation by the hot ISM using
self-consistent three-dimensional hydrodynamical simulations. However, due
to resolution limits, it appears difficult to describe numerically such a
multi-phase gas, even with state-of-the-art hydrodynamical simulations
\citep[e.g.][and references therein]{Kau06,Bry07}. Another useful step
would be to add the results obtained here to semi-analytic models of galaxy
formation \citep{Bow06,Cat06,Cro06}. Of course, a semi-analytic model
is only as good as its input physics and such models are no substitute for
ab-initio modelling from elementary physical principles.

Stripping and ablation of cold gas from galaxies that are orbiting in the
potential well of a system that has abundant virial-temperature gas
undoubtedly plays a large role in determining the morphology of galaxies and
the extension of the red sequence to faint magnitudes. This process is
closely related to the one quantified in this paper.

It now seems very likely that the central factor in dividing
galaxies into the red sequence and the blue cloud is the existence of a
critical halo mass above which gas accumulates at the virial temperature.
What is still unclear is whether the supply of cold gas cuts off in galaxies
associated with massive halos because infalling gas is then shock heated to the
virial temperature \citep{DekB06}, or because cold gas is evaporated and
ablated by the hot gas \citep{Bin04,NipB04}. This paper shows that thermal
evaporation is certainly a non-negligible process.

\section*{Acknowledgments}

We acknowledge helpful discussions with L.~Ferrarese, A.~Graham,
S.~Pellegrini, M.~Sarzi, S.~Tremaine and M.~Volonteri. We thank the
anonymous referee for a constructive report. CN is grateful to Merton
College (Oxford) for hospitality.

\appendix

\section{Evaporation of spheroidal clouds}
\label{appeva}

\subsection{Saturation}
\label{appsat}

The classical heat flux (equation~\ref{eqqcl}) is derived under the
assumption that the mean free path of the electrons is small with
respect to the temperature scale height $T/|\nabla T|$. When this
condition is not satisfied, typically for very high electron
temperature and/or very low electron density, equation~(\ref{eqqcl})
does not hold and the heat flux is said  to be {\it saturated}. As a
discriminant of whether the heat flux in a plasma is saturated,
\cite{CowM77} introduced the parameter $\sigma \equiv \qcl/\qsat$,
where $\qcl=\kappa(\Temp) |{\bf \nabla} \Temp|$ is the modulus of the
classical heat flux, and the modulus of the saturated heat flux is
\begin{equation}
\label{eqqsat}
\qsat=5 f^{1/2} \phisat \rho \cs^3,
\end{equation}
where $\phisat\sim1$ is a dimensionless parameter, $\rho$ is the mass
density of the gas, $\cs=\sqrt{\press/\rho}=\sqrt{\kb \Temp/\mu\mp}$
is the isothermal sound speed, and $\press$ is the pressure. The
quantity $\qsat$ in equation~(\ref{eqqsat}) differs from that in
\cite{CowM77} for the presence of $f^{1/2}$, which we included to
account for the effects of magnetic fields \citep[see][for a
discussion]{CowM77,Bal86}.  The critical value that separates the two
regimes is $\sigma \sim 1$: for $\sigma \ll 1$ the heat flux is
unsaturated, while for $\sigma \gg 1$, equation~(\ref{eqqsat})
applies.  In the case of the evaporation of a cloud modelled as a
prolate or oblate spheroid we get
\begin{equation}
\label{eqsigma}
\sigma=f^{1/2}\sigmaz g(u,v,\epsilon) {\pressism\over\press(u)},
\end{equation}
where 
$\pressism=\lim\limits_{u\to\infty}\press(u)$ is the ISM pressure, 
\begin{eqnarray}
\label{eqsigz}
\sigmaz&=&{2 \kspitzer \Tism^{7/2} \over 25 \phisat \rhoism
  \csism^3 a}\\ &\simeq&{4.5\times 10^{-4}\over\phisat}
\Tismseven^2\neismone^{-1}\akpc^{-1}.\nonumber
\end{eqnarray}
is the saturation parameter\footnote{Our definition of $\sigmaz$
differs from that in \citep{CowS77} by a factor $\cosh\uz$, so that in
the limit $\epsilon \to 0$ we recover the definition of \cite{CowM77}
for a spherical cloud of radius $a$.}  \citep{CowM77}, and $(u,v)$ are
the spheroidal coordinates introduced in Section~\ref{seceva}. Here
$\rhoism$ and $\csism$ are the mass density and isothermal sound speed
of the ISM, and we used $\rhoism\simeq 1.15\mp\neism$ and
$\phisat\sim1$.  The dimensionless function $g(u,v,\epsilon)$ in the
oblate case is
\begin{equation}
\label{eqgobl}
g(u,v,\epsilon)=
-{
\left[\theta(\uz)-\theta(u)\right]^{1/5}
\over
2\left[\theta(\uz)-{\pi\over 4}\right]^{6/5}
\cosh{u}
\sqrt{\sinh^2{u}+\cos^2{v}}
},
\end{equation}
where $\theta(x)\equiv \arctan [\tanh(x/2)]$ , while in the prolate case
\begin{equation}
\label{eqgprol}
g(u,v,\epsilon)=-{\left[\gamma(\uz)-\gamma(u)\right]^{1/5}\over[\gamma(\uz)]^{6/5}\sinh{u}\sqrt{\sinh^2{u}+\sin^2{v}}},
\end{equation}
where $\gamma(x)\equiv\ln [\tanh(x/2)]$, and we recall that
$\uz=\arctanh (1-\epsilon)$.

\begin{figure}
\centerline{\psfig{file=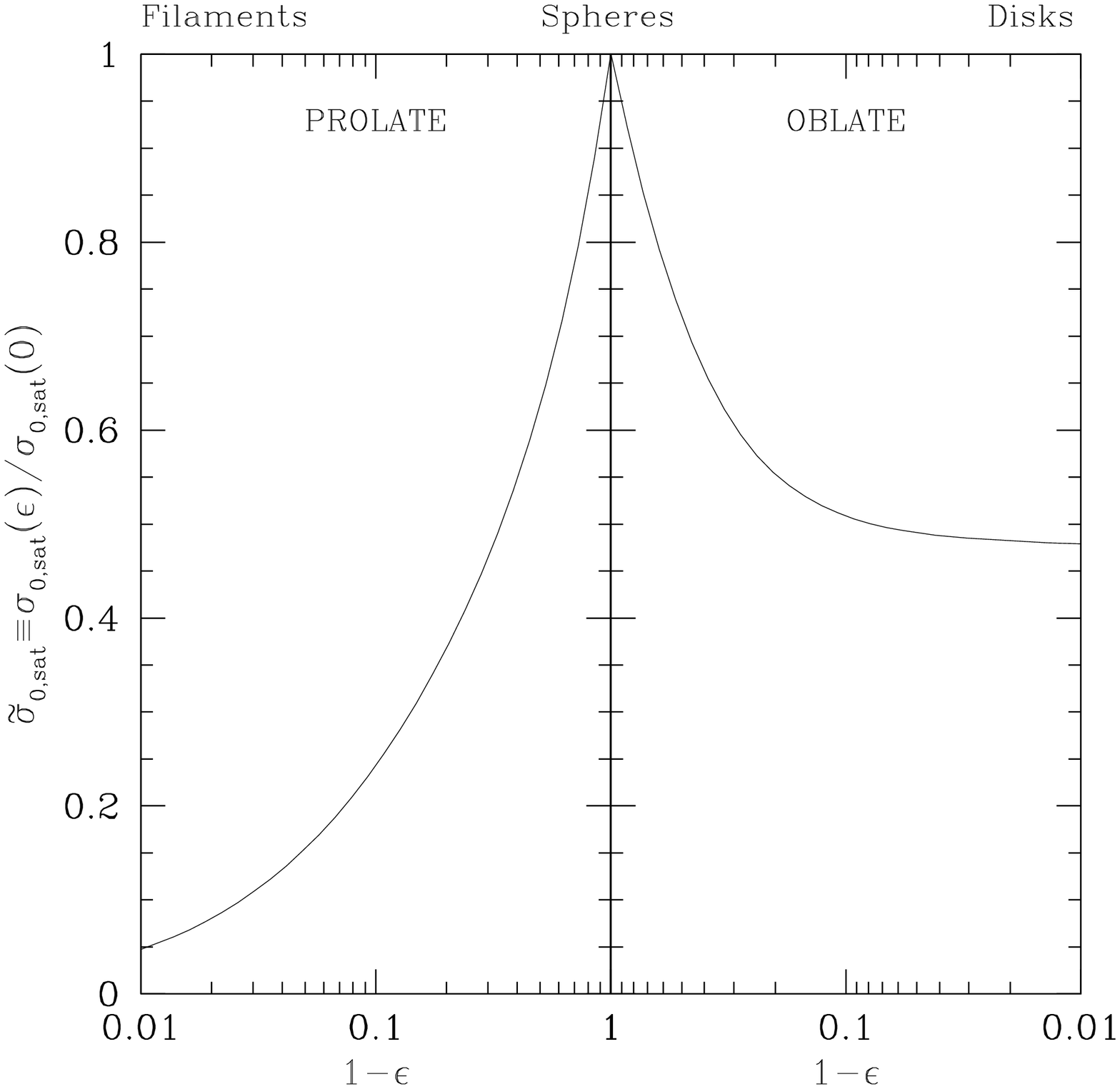,width=0.9\hsize}}
\caption{Top: $\sigmatilzsat$ as a function of the ellipticity $\epsilon$
of prolate (left panel) and oblate (right panel) clouds.}
\label{figsigsat}
\end{figure}


The term $\pressism/\press(u)$, appearing in equation~(\ref{eqsigma}),
is expected to be of the order of unity or smaller, because dense
clouds tend to be overpressured with respect to their environments,
and the pressure is approximately constant throughout the flow if the
Mach number is not large \citep{CowM77}.  While in the case of a
spherical cloud of radius $a$ the condition for saturation
[$\max(\sigma)>1$] is equivalent to $\sigmaz\gsim 1$ \citep{CowM77},
for significantly non-spherical prolate and oblate clouds this
transition occurs at a critical value of $\sigmaz \ll 1$, because the
geometry implies stronger temperature gradients near the equatorial
plane in the oblate case and near the symmetry axis in the prolate
case.  This is reflected in the behaviour of the function
$g(u,v,\epsilon)$ appearing in equation~(\ref{eqsigma}): for $\epsilon
\to 1$ the function $g(u,v,\epsilon)$, at fixed $v$, has a maximum
that significantly exceeds unity, especially for $v \to \pi/2$
(oblate) and $v \to 0$ (prolate). However, even for $\sigmaz$
significantly larger than the critical value, the heat flux is
overestimated only in a small region around $v=\pi/2$ (oblate) or
$v=0$ (prolate), which gives a negligible contribution to the heat
flow.  If the surface of the cloud is $\Stot$, and the surface in
which saturation occurs is $\Ssat$, the fraction of the heat flow that
is overestimated is $\zeta\equiv{\int_{\Ssat} \qclvec\cdot\dSsqvec /
\int_{\Stot} \qclvec\cdot \dSsqvec}$, where $\dSsqvec$ is the area
element.  Integration gives $\zeta=1-\cos\vsat$ (oblate),
$\zeta=\cos\vsat$ (prolate), where $\vsat$ is the value of $v$ at
which the transition between classical and saturated heat flux
occurs. As long as $\zeta \ll 1$, the unsaturated computation still
yields a good approximation of the actual heat flow. So, we define a
reference value of the saturation parameter $\sigmazsat$ (depending on
$\epsilon$) such that $\zeta<1/3$ for
$\sigmaz<\sigmazsat$. $\sigmatilzsat\equiv\sigmazsat(\epsilon)/\sigmazsat(0)$
as a function of $\epsilon$ is plotted in Fig.~\ref{figsigsat}.  The
mass loss rate $\Mdotsat$ for saturated heat flux can be computed
analytically only for spherically symmetric clouds \citep{CowM77},
being
\begin{equation}
\label{eqmdotsat}
\Mdotsat={4 \pi f^{1/2} a^2 \rhoism \csism \phisat F(\sigmaz)},
\end{equation}
with $F(\sigmaz)\simeq2.73\sigmaz^{3/8}$ if $\phisat\sim1$,
where we included the factor $f^{1/2}$ to account for the effects of
magnetic fields.

\subsection{Radiation}
\label{apprad}

\cite{Mck77} showed that a spherically symmetric cold cloud immersed
in a hot ($\Tism\gsim10^5\kelvin$) ambient medium is radiatively
stabilised when $\sigmaz\simeq 0.027/\phisat$.  For given physical
properties of the ambient medium, this condition is satisfied when the
cloud radius is $\arad$ given in equation~(\ref{eqaradsph}), which is
approximately valid also in the oblate case \citep{CowS77}.  In
\cite{NipB04} we computed the critical length $\lcrit$ of radiatively
stabilised filaments of cool gas immersed in a hot medium. So, for
very elongated prolate clouds ($\epsilon \to 1$) an estimate of the
critical size is
 \begin{equation}
\label{eqaradfil}
\arad\simeq {\lcrit\over 2} \simeq 0.07f^{1/2} \lambda^{1/2} \Tismseven^{7/4} \neismone^{-1}{\kpc},
\end{equation}
where $0.08\lsim \lambda(\Tism) \lsim 0.37$ for $10^6 \lsim \Tism/
\kelvin \lsim 10^8$ \citep{NipB04}.  This results has been obtained in
cylindrical symmetry for clouds with electron temperature $\Tc
\sim10^4 \kelvin$, assuming constant pressure in the interface. An
approximated estimate of the evaporation time for cylindrical clouds
of length $l<\lcrit$ is $\tevcyl\sim(l/\lcrit)^2\tcool$, where
$\tcool$ is the cooling time \citep{NipB04}. We verified that in the
temperature range $10^6-10^7 \kelvin$ $\tevcyl\sim\tev$, where $\tev$
is the evaporation time of a cold prolate cloud with
$\epsilon\sim0.98$, and equations~(\ref{eqaradfil}) for filaments and
(\ref{eqaradsph}) for spheres yield to values of $\arad$ within a
factor of two. This suggests that we can safely assume that
equation~(\ref{eqaradsph}) gives an approximated estimate of $\arad$
also for prolate clouds.

\section{Drag force}
\label{appdrag}

Beside gravity, a cloud of cold gas infalling with
velocity ${\bf v}$ through a galaxy's ISM experiences the drag force
\begin{equation}
{\bf F}_{\rm drag}=-{1 \over 2}C S \rhoism v^2 {{\bf v}\over v},
\end{equation}
where $S$ is the cloud cross-section, $\rhoism$ is the density of
the ambient medium at the position of the cloud, and $C$ is the drag
coefficient, which depends on the Reynolds number $R$
\citep[e.g.][]{Lan59}. In our applications $R \ll 10^5$, and we
crudely assume $C(R)=24/R$ for $R<48$ and $C(R)=0.5$ for $R\geq 48$.

To estimate the effect of the drag force on the dynamics of the clouds
we consider here the simple case of a spherical cloud of mass $M$
infalling radially in the model galaxies  LM and HM described in
Section~\ref{secmod}. Integrating the equation of motion
\begin{equation}
\label{eqdrag}
\ddot{r}=-{G \Mdyn(r)\over r^2}+{1 \over 2 M}C(r) S(r) \rhoism(r) \dot{r}^2
\end{equation}
between $r$ and 0 with $v(r)=0$, we obtain a dynamical time
$\tdyn(M,r)$. We note that at
fixed mass, the cross-section depends on $r$ because the cloud
density does (see Section~\ref{seccal}).  The star-formation mass $\Msf$ is
such that $\tev(\Msf,r)=\tdyn(\Msf,r)$.  We found that the values of
$\Mmin=\max_{r>0}\Msf(r)$ obtained considering equation~(\ref{eqdrag})
are larger than those obtained in the absence of the drag force by
$10-15$ per cent in the model galaxies LM and HM.  We conclude that
neglecting the effect of drag implies only a modest underestimation of
the value of the minimum mass $\Mmin$.

\bsp

\label{lastpage}

\end{document}